\documentclass[12pt]{article}
\usepackage{graphicx}
\begin{document}

\title{Longitudinal tension and mechanical stability of a pressurized straw tube}

\author{L.~Glonti, T.~Enik, V.~Kekelidze, A.~Kolesnikov, D.~Madigozhin\thanks{Corresponding author. Email address: madigo@mail.cern.ch}, \\ 
N.~Molokanova, S.~Movchan, Yu.~Potrebenikov, S.~Shkarovskiy \\
Joint Institute for Nuclear Research, 141980, Dubna, Russia}

\maketitle

\begin{abstract}
For the development of charged particle detectors based on straw tubes operating in vacuum, 
a special measurement technique is required for the evaluation of their mechanical properties. 
A summary of the known equations that govern straw behavior under internal pressure 
is provided, and a new experimental method of a strained pressurized straw tube study is 
presented in this paper. The Poisson's ratio of the straw wall, which defines the stability 
conditions of a built-in tube, is measured for the NA62 spectrometer straw, and its minimum 
pre-tension is estimated.
\end{abstract}

keywords: straw tracker, straw tube, vacuum, pressure 

PACS[2010] 29.40.Cs 29.40.Gx

\section{Introduction}

During the past few decades, a series of experiments in high-energy physics have been 
designed to investigate very rare decay modes of kaon (see, for example, \cite{NA62TD,Buonaura:2016gho}). 
In particular, the purpose of NA62 experiment at CERN SPS \cite{NA62TD} is to measure 
$K^+ \to \pi^+ \nu \bar \nu$ decay branching ratio of the order of $10^{-10}$ with an 
uncertainty of 10\%. It will be a test of the Standard Model and a probe to possible
new physics. 
This is a challenging task, requiring an unprecedented precision of $\pi^+$ momentum 
measurement in order to reject the dominating background using the evaluation of event 
missing mass. This require a precise measurement of the momentum and position of the 
charged particle. 

A minimal material budget requirement in such experiments stimulates the development of 
gaseous particle detectors based on straw tubes containing gas under atmospheric pressure 
and operating in vacuum. These tubes maintain an internal overpressure of 1 atm, but 
higher overpressure may be desirable for a slower drift of electrons in the gas.

From the mechanics of a column, it is known that, if a compressive axial load is applied 
to a clamped thin-walled tube without overpressure, the load critical value $F_{crit}$ 
for buckling is 
\cite{Timoshenko1961,Catinaccio2010}
\begin{equation}
F_{crit} = 4 \pi^2 E J/L^2.
\label{fcrit}
\end{equation}
where $R$ is the straw radius, $L$ is its length, and $h$ is the thickness of the tube wall.
Further, $E$ is the Young's modulus and $J = \int \int y^2 dx dy = \pi R^3 h$ is a second moment of the cross-section of the 
tube wall . $EJ$ is known as the object flexural rigidity. For an NA62 straw tube 
$F_{crit} \approx 0.5 N$, which is a small value.

It should be considered that a large straw deviation 
may appear when this limit is approached \cite{Catinaccio2010}. A shift of a few millimeters of the tube axis with respect to the 
anode wire may cause an electric discharge and detector malfunction. Only axial tension 
guarantees the buckling prevention for a long straw tube, and it has been demonstrated earlier 
\cite{Catinaccio2010,Wertelaers2010} that an internal overpressure changes this necessary 
minimum tension.

Moreover, pre-tension is required to maintain the curvature caused by gravity to be small for 
a straw placed horizontally. For the given straw geometry and overpressure, the only parameter 
for the curvature control is the preliminary stretching force applied to the straw.
 
Therefore, estimation of the minimum pre-tension of the straw is necessary for any design of a straw tracker 
operating in vacuum. However, the calculations based on the published material properties of the straw wall 
are not sufficiently precise, because these properties depend on the batch of material. Therefore, an experimental 
procedure for the evaluation of material properties of a straw under operating conditions may be very useful.

The main purpose of this article is to present a new experimental method to estimate the 
minimum preliminary stretching force required to prevent the pressurized straw buckling and
to limit its gravitational curvature.

\section{Mechanics of a pressurized straw tube}

The pressurized tube problem is not new in the industry, but it is not well-known in high-energy 
physics instrumentation. Therefore, we provide a summary of the known equations 
driving the behavior of straw tubes under internal pressure.

Only two stress directions are essential for the problem of a strained straw under internal 
pressure: the axial direction along the straw axis ($\parallel$) and the circumferential 
direction ($\perp$) tangent to the cylindrical surface of the straw in the plane normal to the axis. 
For a thin-walled tube ($h \ll R$), it can be demonstrated that straw circumferential (hoop) 
stress caused by the inner overpressure $P$ is always $\sigma_\perp = \frac{PR}{h}$ \cite{Gere2004}. 
The axial stress $\sigma_\parallel$ depends on the conditions at the straw ends.

\subsection{Pressurized tube with free closed ends}

We will begin with the simple formulae derived in the isotropy approximation, when the material properties 
do not depend on the considered direction. If the internal pressure acts not only on the cylindrical shell 
of the tube but also on the free ends closed by airtight end plugs, a simple relation can be derived between the hoop stress $\sigma_\perp$ and the free closed tube axial stress 
$\sigma^{free}_\parallel = \frac{P \pi R^2}{2\pi R h} = \frac{PR}{2h} = \sigma_\perp /2$ 
(see \cite{Gere2004} for cylindrical pressure vessel).

Hoop stress leads to the transverse strain of the material of the straw wall defining the direct 
contribution to the relative radius increasing $\epsilon^d_\perp = \sigma_\perp / E = \frac{PR}{Eh}$.
For the typical straw tube $\epsilon^d_\perp \ll 1$; thus, this value may be used as a small 
parameter to determine the order of the considered contribution to a strain. The direct contribution 
to the axial relative elongation $\epsilon_\parallel = \frac{\Delta L}{L}$ for the free closed tube has 
the same first order: $\epsilon^d_\parallel = \sigma^{free}_\parallel / E = \frac{PR}{2Eh}$.

The Poisson effect leads to the shortening of the straw wall in the direction perpendicular to the corresponding direct strain. 
The Poisson ratio $\mu$ is defined as the ratio of the perpendicular relative shortening to the corresponding 
direct relative elongation. Thus, we have the first approximation for a free closed tube.
\begin{eqnarray}
\label{strainlfree}
\frac{\Delta L}{L} = \epsilon_\parallel  = \epsilon^d_\parallel - \mu \epsilon^d_\perp = \frac{PR}{2hE}(1-2\mu) \\
\frac{\Delta R}{R} = \epsilon_\perp     = \epsilon^d_\perp - \mu \epsilon^d_\parallel = \frac{PR}{2hE}(2-\mu).
\label{straintfree}
\end{eqnarray}
In the paper \cite{Peshekhonov}, the equation (\ref{strainlfree}) has been observed to be sufficiently precise 
to the best of our knowledge of the material properties. The hoop strain formula (\ref{straintfree}) has been 
derived for the pressure vessel in \cite{Shells}.

In an experimental set-up employing straws in vacuum, straw ends are usually built into the rigid 
frame (built-in tube) \cite{NA62TD}. For this case, the formulae (\ref{strainlfree}, \ref{straintfree}) 
in general are incorrect, since the pressure force applied to the built-in end is balanced by the rigid frame
rather than by the axial tension of the wall.

However, there is a special case wherein a built-in tube and a free closed tube are equivalent.
Further, it can be used to obtain the simplest estimation of the minimum preliminary tension required 
to prevent straw buckling in vacuum \cite{DeCarloLivio}. If a free closed pressurized straw 
is glued into a rigid frame that exactly fits the straw length enlarged by the overpressure 
(\ref{strainlfree}), the frame does not apply any axial load to the built-in straw. If the
frame requires a longer straw, an overall axial tension is applied to the straw ends ensuring the 
stability of the tube against buckling.
Furthermore, no change occurs in the state of the straw wall if we 
connect the inner straw volume with the gas supply mounted in the frame.

Therefore, in order to prevent buckling, we require a minimum straw pre-tension ensuring the 
elongation (\ref{strainlfree}) prior to the vacuum creation around the tube.
The stress caused by straw preliminary tension $T_0^{min}$ should be at least 
$\frac{T_0^{min}}{2\pi Rh} = E \epsilon_\parallel =  \frac{PR}{2h}(1-2\mu)$, and the 
minimum pre-tension for buckling prevention is 
\begin{equation}
T_0^{min} = P \pi R^2 (1-2\mu).
\label{mintension}
\end{equation}

\subsection{Pressurized tube with built-in ends}

We will assume the different material properties in the axial and circumferential directions 
(orthotropic approximation), which is quite usual for a straw wall material. For example, NA62 straws 
are made of Hostaphan\textsuperscript{\textregistered} polyethylene terephthalate (PET) film. The film manufacturer reports the 
different values of transverse (transverse direction, TD) and longitudinal (machine direction, MD) 
Young's moduli \cite{Hostaphan} (see Table \ref{strawprop}). Therefore, the direct contribution to the 
relative radius change of straw becomes $\epsilon^d_\perp = \sigma_\perp/E_\perp = PR/(E_\perp h)$.

Owing to the Poisson effect, the hoop strain causes an axial strain of the opposite sign, and thus, the straw 
becomes shorter if the ends are not built-in. This imaginary relative change of the length is 
$\epsilon^\mu_\parallel = -\mu_\perp \epsilon^d_\perp$, where $\mu_\perp$ is the transverse (circumferential
for the tube) Poisson's ratio. However, the straw ends are fixed, which indicates an appearance of the compensatory 
tension force $\Delta T$ and the corresponding axial stress $\Delta T/(2\pi Rh)$, which returns the straw to 
its initial length:
\begin{equation} 
\frac{\Delta T}{2\pi Rh} = - E_\parallel \epsilon^\mu_\parallel = \mu_\perp P\frac{R}{h} \frac{E_\parallel}{E_\perp}.
\end{equation}

From Maxwell's reciprocity theorem, the known relation 
$\mu_\parallel = \mu_\perp \frac{E_\parallel}{E_\perp}$ can be derived, where $\mu_\parallel$ is the axial 
Poisson's ratio \cite{Catinaccio2010}. Thus, the wall axial tension is 
\begin{equation}
T = T_0 + \Delta T = T_0 + 2 \mu_\parallel P \pi R^2 .
\label{tformula}
\end{equation}

The total force applied by a gas-filled straw in vacuum on the detector frame can be considered as an 
``effective tension'' $T_P = T - P\pi R^2$, if the effect of external atmospheric pressure applied to 
the frame is calculated by ignoring all holes made for the gas supply into the straws. For this 
effective tension, we obtain
\begin{equation}
T_P = T_0 - (1 - 2\mu_\parallel) P \pi R^2.
\label{tpformula}
\end{equation}
Poisson's ratio $\mu$ for plastics is usually less than 0.5. Thus, the vacuum around a straw 
leading to its inner overpressure diminishes $T_P$ with respect to the straw pre-tension $T_0$ 
in spite of the increase in true straw tension $T$ (\ref{tformula}).

\subsection{Lateral effect of internal pressure}

\begin{figure}[t]
\begin{center}
\resizebox{0.6 \columnwidth}{!}{%
\setlength{\unitlength}{1mm}
\begin{picture}(160,80)
\includegraphics[width=160mm]{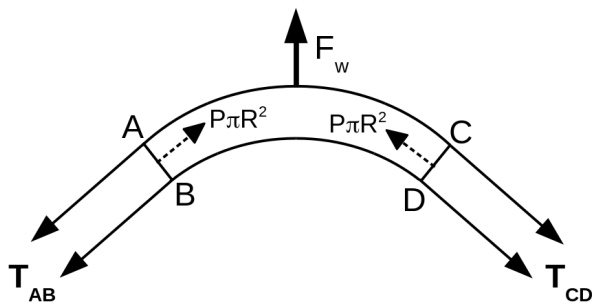}
\end{picture}
}\end{center}
\caption{Forces applied to the curved tube element without flexural rigidity subject to internal 
pressure and axial tension}
\label{fig:transforce}
\end{figure}

We will evaluate the summary force applied by internal pressure to the wall of a curved tube. 
Ignoring the flexural rigidity, consider a short element of a slightly curved tube (Figure \ref{fig:transforce}), 
limited by its two cross-sections normal to the curved axis. The absolute longitudinal tension force 
$|T_{AB}|=|T_{CD}|=|T|$ is a constant in this approximation, but the directions of $T_{AB}$ and $T_{CD}$ 
forces are defined by different slopes at the element ends.

The area of the arched side of the wall (upper side in Figure \ref{fig:transforce}) is larger 
than the area of the concave part. Consequently, a non-zero summary pressure force $F_W$ applied to 
the wall appears, which is directed toward the arched side of the wall. Thus, the internal overpressure 
always attempts to increase the existing curvature of the tube.

The evaluation of summary pressure effect has been performed earlier in a few ways: an imaginary ideal 
piston at the boundary of the pressurized tube segment \cite{Haringx1952}, the consideration 
of the forces applied to separated fluid contained within a tube element \cite{Palmer1974}, 
and a direct integration of the pressure forces \cite{Catinaccio2010}. 

However, we can directly evaluate the pressure effect $F_W$ without integration. 
For the given $P$, the summary force $F_W$ depends only on the shape of the tube segment. Imagine 
an absolutely rigid shell consisting of a curved tube wall with additional transverse 
plugs closing the cross sections AB and CD tightly. A resultant force applied by inner overpressure 
to a closed rigid shell is always zero. Therefore, the total pressure force applied to the 
curved wall is equal, but with the opposite sign, to the vector sum of the pressure forces $P\pi R^2$ 
applied to the end plugs, which are perpendicular to the curved axis of the tube at the ends 
(see Figure \ref{fig:transforce}). One of these forces is directed against $T_{AB}$, and the 
other against $T_{CD}$, which effectively diminishes the straw tension by the 
pressure-related force $P\pi R^2$.

Thus, the lateral dynamics of a curved tube element are defined by the effective tension 
$T_P = T - P\pi R^2 = T_0 - (1 - 2\mu_\parallel) P \pi R^2$, which replaces the true tension 
$T$ in all the formulae describing a straw bending \cite{Palmer1974,Catinaccio2010,Wertelaers2010}.

It is important to understand the physical difference between the effective tension $T_P$ 
defining the straw bending phenomena and the tube wall tension $T$ in the tensile strength 
formulae. Straw wall stress is normally increased owing to the internal overpressure, whereas the pressurized 
straw as an elastic body is effectively relaxed under the same conditions, and this relaxation is described by the
pressure-dependent effective tension $T_P$ behavior.

For $T_P < 0$, a summary transverse force pushes a curved straw element toward its arched 
side, and without flexural rigidity, the curvature increases until the tension (increased 
owing to the curved tube elongation) becomes equal to the pressure force everywhere along 
the tube: $T = P\pi R^2$.

From $T_P > 0$ and (\ref{tpformula}), we obtain 
the minimum longitudinal straw pre-tension $T_0$ to prevent buckling \cite{Catinaccio2010}
\begin{equation}
T_0^{min} = (1-2\mu_\parallel)P\pi R^2. 
\label{buckling}
\end{equation}
The same limit is defined by (\ref{mintension}) derived in a different way.

\subsection{Equilibrium of a horizontal straw}

For a horizontally placed straw, we should ensure that the 
maximum vertical deviation (sagitta $Sag$) caused by gravitation is small.
Consider a horizontal straw with built-in ends subject to vertically distributed 
gravitational load $q$ and an internal overpressure $P$. We choose the origin of the 
axial coordinate $x$ at the center of the straw. The axis of a straw lateral deviation 
$y$ is directed downwards, thus $y(0)=Sag$. The linear gravitational load is $q = g \rho$, 
where $\rho$ is the linear density of the straw and $g$ is the gravitational acceleration. 

The straw equilibrium equation for this case was derived in \cite{Catinaccio2010,Wertelaers2010}:
\begin{equation}
\frac{dF}{dx} = - EJ\frac{d^4 y}{dx^4} + T_P \frac{d^2y}{dx^2} + q = 0,
\label{equilibrium}
\end{equation}
where $\frac{dF}{dx}$ is a resulting force per unit of length, which is zero for the 
case of static equilibrium.

Straw ends in the NA62 spectrometer are glued into the frame (clamped). Therefore, the boundary 
conditions are $\frac{dy}{dx}(\pm \frac{L}{2}) = 0$ and $y(\pm \frac{L}{2})=0$.
The symmetric solution ($y(x)=y(-x)$) for this static case is
\begin{equation}
y_s(x) = \frac{q}{2 T_P} ( \frac{L^2}{4} - x^2 +\frac{L}{k} \cdot \frac{cosh(kx)-cosh(kL/2)}{sinh(kL/2)} ), \label{static} 
\end{equation}
where $k=\sqrt{\frac{T_P}{EJ}}$ (see \cite{Wertelaers2010}). Sagitta $y_s(0)$ is 
increased by the internal overpressure owing to the decrease in $T_P$ (\ref{tpformula}).

\subsection{Low-frequency oscillations}

The interesting consequence of the pressure lateral effect is a straw vibration
pressure dependence caused by the behavior of $T_P$ used instead of string tension 
for straw transverse dynamics.
For the case of oscillations, we should set $\frac{dF}{dx} = \rho \frac{d^2y}{dt^2}$ instead 
of zero in (\ref{equilibrium}). We will determine the solution in the form of 
$y(x,t) = y_s(x)+y_f(x,t)$, where $y_s(x)$ is a solution of the 
static equation (\ref{equilibrium}). It leads to the wave equation
\begin{equation}
\rho \frac{d^2 y_f}{dt^2} = - EJ\frac{d^4 y_f}{dx^4} + T_P \frac{d^2y_f}{dx^2}
\label{wave}
\end{equation}
set up by Lord Rayleigh \cite{Rayleigh1894}.

We will determine the symmetric solution with fixed ends ($y_f(\pm \frac{L}{2})=0$) 
and with ``pinned'' boundary conditions ($\frac{d^2y_f}{dx^2}(\pm \frac{L}{2}) = 0$). 
Such a simple solution is sufficient for the qualitative understanding of straw oscillations. 
These boundary conditions are satisfied for $y_f(x,t) = cos((1+2n)\pi x/L)e^{i 2 \pi f t}$ 
with any integer $n$. For $n=0$, the equation (\ref{wave}) results in $(2 \pi f)^2 \rho =  (\pi/L)^2 (EJ (\pi/L)^2 + T_P)$, 
and thus the lowest frequency of the straw vibration is \cite{Fletcher1964}
\begin{equation}
f  =  \frac{1}{2L}\sqrt{ \frac{EJ (\pi/L)^2 + T_P}{\rho}}.
\label{frequency}
\end{equation}
Notably, the flexural rigidity of the straw results in an addition of $EJ (\pi/L)^2$ to the 
effective tension in (\ref{frequency}). The frequency is decreased by the overpressure 
owing to the behavior of $T_P$ (\ref{tpformula}).

\subsection{Radius correction}

$T_P$ depends on the straw radius $R$ considered so far as a constant value. However, $R$ depends on $P$, 
whereas the radius of the built-in end plug remains unchanged. It leads to the formation of a radius transition zone 
on the tube wall near the straw end. However, the axial component of the pressure force applied to the transition 
zone is subtracted from the axial wall tension applied to the end plug. Therefore, for $T_P$ calculation, 
we can consider the radius of the end plug to be equal to the pressure-dependent straw radius $R$.

The direct contribution of overpressure to the relative change of radius is $\frac{\sigma_\perp }{E_\perp} = \frac{PR}{E_\perp h}$, 
but if we consider the additional tension applied to the built-in ends, this term becomes 
$(1 - \mu_\parallel \mu_\perp)\frac{PR}{E_\perp h}$. Moreover, the 
applied pre-tension $T_0$ diminishes the straw radius owing to the Poisson's effect. Therefore, we have
\begin{equation}
R = R_0 (1 + (1 - \mu_\parallel \mu_\perp)\frac{PR_0}{E_\perp h} - \frac{\mu_\parallel T_0}{E_\parallel 2\pi R_0 h}),
\label{radius}
\end{equation}
where $R_0$ is the initial radius of the straw. For $T_P$ calculations, if $T_0 \approx 1$~kgf, the radius change 
(\ref{radius}) leads to the next-order correction in terms of the small parameter $\frac{PR}{Eh}$.

\section{Poisson's ratio measurement}

It can be observed from (\ref{buckling}) that the straw buckling limit for a given overpressure is defined 
by the tube radius $R$ and the Poisson's ratio $\mu_\parallel$. The Poisson's ratio is typically not 
provided in the PET film specifications. The published independent measurements of $\mu$ are not related 
to the specific batch of material used in the straw tubes production for the given detector. Moreover, the straw production 
process may change some foil properties. Therefore, the $\mu_\parallel$ measurement procedure applicable to a 
welded straw is required in order to evaluate the minimum straw pre-tension for the specific detector design.

\subsection{Straw specimens}

Two straw specimens have been tested (see Table \ref{strawprop}). The tubes are made of 
PET Hostaphan\textsuperscript{\textregistered} foils using longitudinal welding
\cite{NA62TD}. The parameters of the foils can be found in the manufacturer specifications \cite{Hostaphan}.

\begin{table}[!h]
\caption{Properties of the tested straws}
\begin{center}
\begin{tabular}{|l|l|l|} \hline
            Property                      &  9.8-mm straw & 18-mm straw  \\ \hline 
Diameter,  mm                             &  9.8(1)      & 18.0(1)     \\ 
Length,    m                              &  2.10(1)     & 1.90(1)     \\ 
Material density, g/cm$^3$                &  1.4(1)      & 1.4(1)      \\
Wall thickness, $\mu$m                    &  36(1)       & 54.4(8)     \\  
Tube linear density, g/m                  &  1.55(12)    & 4.31(6)     \\  
$E_\parallel$, N/mm$^2$                   &  4500(500)   & 4000(500)   \\ 
$E_\perp$,     N/mm$^2$                   &  5000(500)   & 5500(500)   \\ \hline
\end{tabular}
\end{center}
\label{strawprop}
\end{table}

The first specimen (9.8-mm straw) is obtained from a party of approximately 7000 straws produced at the Joint Institute 
for Nuclear Research (Dubna) for NA62 spectrometer \cite{NA62TD}. This 9.8-mm straw is coated inside the tube with 
two thin metal layers ($0.05 \mu m$ of Cu and $0.02 \mu m$ of Au) in order to provide electrical conductivity on the 
cathode and to improve the impermeability of the straw tube. The wall material volume density \cite{Hostaphan} is used to 
estimate the linear density of the 9.8-mm straw. The contributions of metal layers and the air mass inside the tube to linear density are negligible.

The second specimen (18-mm straw) is made of a thicker Hostaphan\textsuperscript{\textregistered} film. In 
this case, the linear density of the tube has been measured by weighing, since the wall thickness is not strictly defined in the
manufacturer specifications.

\subsection{Test bench for studies of a strained straw under pressure}

\begin{figure}[h]
\begin{center}
\resizebox{1.0 \columnwidth}{!}{%
\setlength{\unitlength}{1mm}
\begin{picture}(160,45)
\includegraphics[width=160mm]{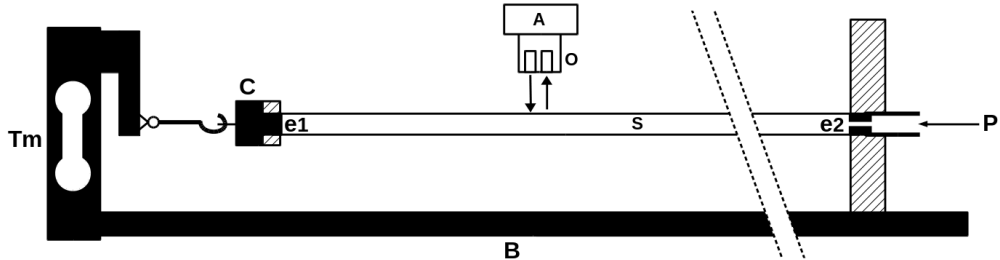}
\end{picture}
}\end{center}
\caption{The test bench scheme: 
B -- rigid basement; Tm -- tensometer; e1 -- closed end plug; e2 -- end plug with a gas supply channel; C -- end cap with a seal; 
P -- pressure supply; S -- straw; O -- optical coupler; A - amplifier.}
\label{fig:testbench}
\end{figure}

A special test bench (see Figure \ref{fig:testbench}) has been manufactured in order to study the properties
of a built-in straw with an initial pre-tension and an inner overpressure applied subsequently. The 
longitudinal force applied to the straw end $T_P = T - P \pi R^2$ is measured using a tensometer $Tm$ based on 
a single-point aluminum load cell (Tedea-Huntleigh, model 1022) \cite{tensometer}.

A straw specimen is closed on the tensometer side by the end plug $e1$ with the end cap $C$ using glue for rigid sealing. 
The end cap $C$ is connected to the tensometer $Tm$ using two flexible joints and a rigid rod. The other straw end 
contains a plug $e2$ with a gas supply channel. This straw end is glued into the solid support, which may be moved along 
the rigid basement $B$ and fixed at a specific place in order to create a preliminary straw tension prior to the 
test. Subsequently, the pressure supply is opened, and the changing pressure values $P$ are recorded together with the corresponding 
results of the tensometer measurement $T_P$ (see points in Figure \ref{fig:vspred2} and Figure \ref{fig:coplot}A).

Straw oscillations are studied using the optical coupler O \cite{NA62TD}, which emits 
constant intensity infrared radiation and registers the radiation reflected from the straw wall. When the 
straw oscillations are mechanically excited, the registered radiation intensity is modulated by the changing 
distance to the straw wall. The obtained signal is amplified and sent to the oscilloscope with a fast Fourier transform 
function. The lowest frequency peak position is registered as the lowest frequency of the straw (points in Figure \ref{fig:coplot}B).

The end cap $C$ is slightly shifted down owing to its weight of $10-20$~gf. However, for the horizontal 
force $T_P > 300$~gf (assuming the cap weight of $25$~gf), the vertical shift of the cup is less than $4\ mm$. 
It leads to the relative elongation of 2-m straw $\approx 10^{-4}$, which is much less than 
the minimum straw elongation caused by the tested preliminary tension ($10^{-3}$). Thus, the straw ends 
may be considered to be fixed during the test.

\subsection{Effective tension and Poisson's ratio measurement}

It is known that any material becomes nonlinear for a large stress, whereas the linear
properties of the material are defined for zero-stress limit. However, on the built test bench, precision measurement 
becomes problematic for a low effective tension $T_P$. Therefore, we must consider the possible 
nonlinearity in such a way that the resulting $\mu_\parallel$ could be easily extracted for $T \to 0$.

Accordingly, we postulate a weak linear dependence of Poisson's ratio as a function 
of axial stress. For the fits of our experimental results, we use the tension-dependent value 
\begin{equation} 
\nu = \mu_\parallel - \frac{kT}{2\pi R h}
\label{nuform}
\end{equation}
instead of only $\mu_\parallel$. Apart from the material nonlinearity, the coefficient $k$ also absorbs 
the effect of the set-up deformation under tension and the next-order effects ignored in the 
formulae. Expression (\ref{nuform}) provides a physically motivated interpolation and extrapolation 
of the measurement results, whereas in order to compare the resulting Poisson's ratio with the other 
measurements, we can consider the measured $\mu_\parallel$ and ignore the stress-dependent term.

$T$ and $R$ variables in (\ref{nuform}) depend on the Poisson's ratio value. Therefore, we 
implement an iterative procedure starting with a tension $T=T_0$, nominal straw radius $R=R_0$, 
and the starting Poisson's ratio value of $\nu = \mu_\parallel$. In each iteration, new $T,\nu,R$ 
values were calculated, and three iterations were sufficient for the precise calculation.

\begin{figure}[h]
\begin{center}
\resizebox{0.7 \columnwidth}{!}{%
\setlength{\unitlength}{1mm}
\begin{picture}(100,100)
\includegraphics[width=100mm]{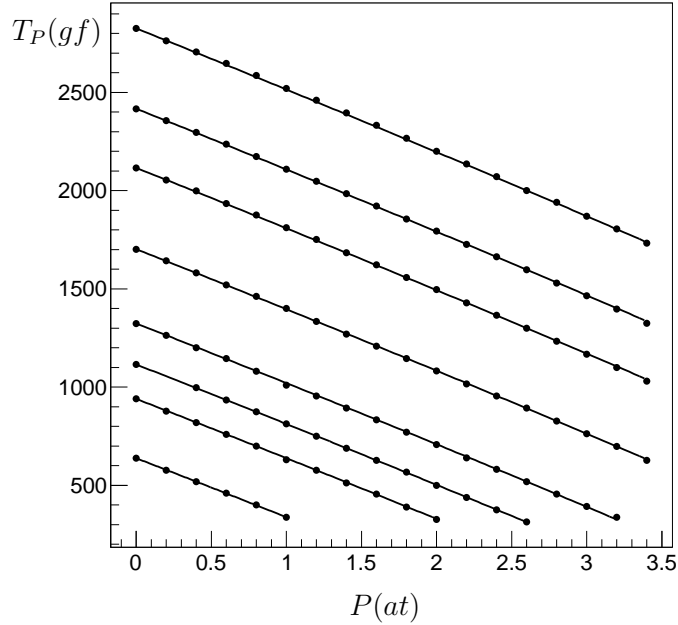}
\put(-56,0){$P (at)$}
\put(-104,82){$T_P (gf)$}
\end{picture}
}\end{center}
\caption{9.8-mm straw effective tension $T_P$ versus the overpressure $P$ for 
different initial tensions $T_P(0)$. Circles -- measurements; solid lines -- fit by 
the formula (\ref{tpfitform}).}
\label{fig:vspred2}
\end{figure}

Two free parameters ($\mu_\parallel$ and $k$) describe the measured tensions and pressures satisfactorily. 
Figure \ref{fig:vspred2} and Figure \ref{fig:coplot}A show the measured effective tensions 
for the NA62 straw along with the result of their fit with the formula 
\begin{equation}
T_P = T_0 - (1-2\nu) P \pi R^2.
\label{tpfitform}
\end{equation}
The MINUIT \cite{James:1975dr} package and ROOT \cite{Brun:1997pa} interface were used to obtain the 
resulting parameter values and their fit errors.

\begin{table}[!h]
\caption{Poisson's ratio measurement results}
\begin{center}
\begin{tabular}{|l|l|l|l|l|l|l|} \hline
                                     &     \multicolumn{3}{c|}{9.8~mm straw}   &  \multicolumn{3}{c|}{18~mm straw}        \\ \hline 
                                     & $k$, $\frac{\mu m^2}{gf}$  &  $\mu_\parallel$ &  $\mu_\perp$  & $k$, $\frac{\mu m^2}{gf}$ & $\mu_\parallel$ & $\mu_\perp$  \\ \hline 
Central value                        & 3.98 &  0.3055          &  0.3394       & -2.89   & 0.2960          & 0.4070       \\ \hline 
Radius correction                    & 0.43 &  0.0005          &  0.0006       &  0.92   & 0.0005          & 0.0007       \\ 
Radius value                         & 0.14 &  0.0083          &  0.0092       &  0.59   & 0.0039          & 0.0054       \\ 
$T_P$ scale                          & 0.06 &  0.0099          &  0.0110       &  1.24   & 0.0088          & 0.0121       \\ 
$P$ scale                            & 0.27 &  0.0094          &  0.0104       &  1.03   & 0.0084          & 0.0116       \\ 
$\delta \frac{E_\perp}{E_\parallel}$   & 0    &  0               &  0.0306       &  0      & 0               & 0.0296       \\ \hline
Systematic error                     & 0.53 &  0.0160          &  0.0353       &  1.95   & 0.0128          & 0.0344       \\ \hline
Statistical error                    & 0.18 &  0.0005          &  0.0006       & 14.35   & 0.0150          & 0.0206       \\ \hline 
Total error                          & 0.56 &  0.0160          &  0.0353       & 14.48   & 0.0197          & 0.0401       \\ \hline 
\end{tabular}
\end{center}
\label{fitresults}
\end{table}

The fit to all the measured $T_P$ values for the given straw is performed with a common set of 
free parameters and the same assumed measurement error. The error ($\pm 3.6$~gf for the 
9.8-mm straw and $\pm 20.47$~gf for the 18-mm straw) is defined in such a way that the resulting 
$\chi^2 / ndf =1$, in order to estimate the parameter statistical uncertainties.

The fit results and systematic uncertainty contributions are shown in Table \ref{fitresults}.
The coefficients of correlation between $\mu_\parallel$ and $k$ are 0.991 for the 9.8-mm straw and 1.000 
for the 18-mm straw. Thus, for the second specimen, the presence of the non-zero term $k$ is not confirmed.

The radius correction effect is considered, but it is also completely included in the 
systematic errors as a ``Radius correction'' contribution. 
The Gaussian width of the NA62 straw diameter distribution is approximately 0.03 mm, and the systematic 
shift of its central value from the nominal number has the same size \cite{NA62TD}. Thus, considering a  
conservative value of the possible diameter systematic uncertainty of 0.1 mm for the tested specimens, 
we have obtained the ``Radius value'' contribution to the systematic error shown in Table \ref{fitresults}.
Moreover, the effects of a systematic scale shift of 5\% on both the measured effective 
tension and measured pressure are considered as independent contributions (``$T_P$ scale'' and ``$P$ scale'') 
to the systematic errors.

Considering the manufacturer’s information about Young's moduli (see Table \ref{strawprop}),
the $\mu_\perp$ central values shown in Table \ref{fitresults} have been extracted as 
$\mu_\perp = \mu_\parallel \frac{E_\perp}{E_\parallel}$. Their uncertainties depend on the error of the modules 
ratio, which may be approximately estimated from the significant digits of the provided values as 
$\delta \frac{E_\perp}{E_\parallel} \approx 0.1$.

Typically, the reported Poisson's ratio values for the oriented PET films are 0.37-0.44 \cite{PET}, 
and thus, the obtained $\mu_\perp$ values are consistent with them. However, the resulting $\mu_\parallel$ is approximately $0.3$, 
which may be a specific property of Hostaphan\textsuperscript{\textregistered} foil or the consequence 
of the production of straw tube using longitudinal welding \cite{NA62TD}.

\section{Oscillation frequency measurements}

The oscillation frequency measurement was the last test of the straw tensions performed 
on the assembled NA62 spectrometer modules \cite{NA62TD}. The same test has been 
repeated on the present test bench for a qualitative verification of the pressure effect.

The length of the tube on the test bench is defined by the position of the movable 
support and the end cup $C$ position. Unfortunately, when a straw tension is applied 
to the end cup fixed on the short rod with flexible joints, an effective oscillator is 
formed with a frequency close to the measured frequency of straw. Therefore, using this 
test bench, only a qualitative understanding of straw oscillations can be obtained using the simple frequency formula (\ref{frequency}).	
	
\begin{figure}[t]
\begin{center}
\resizebox{0.8 \columnwidth}{!}{%
\setlength{\unitlength}{1mm}
\begin{picture}(160,160)
\includegraphics[width=160mm]{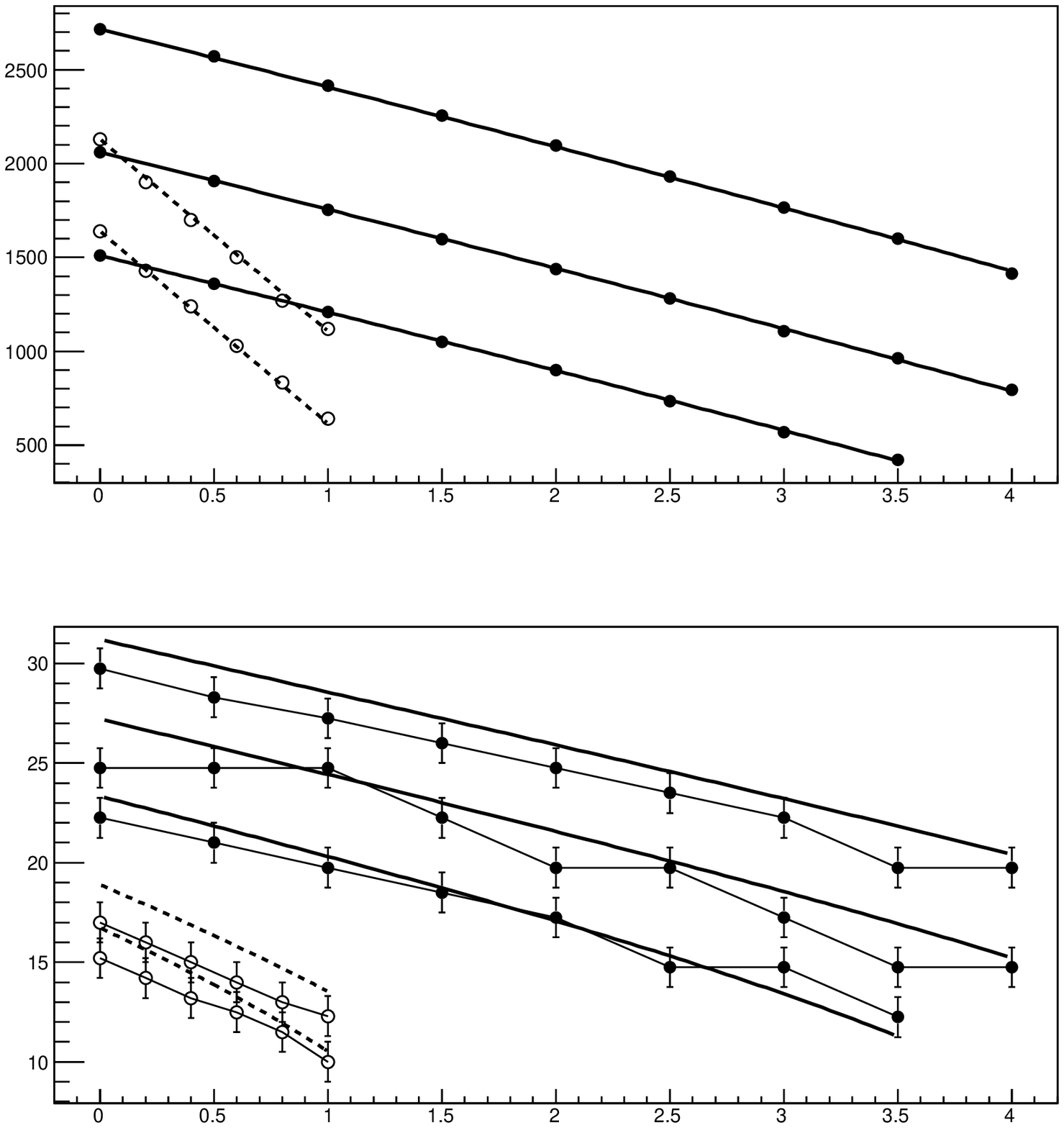}
\put(-80,0){$P (at)$}
\put(-162,120){$T_P (gf)$}
\put(-162,40){$f (Hz)$}
\put(-32,140){A}
\put(-32,60){B}
\end{picture}
}\end{center}
\caption{A: effective tensions $T_P$ versus $P$ for different initial tensions $T_P(0)$. 
B: lowest frequency for the corresponding $P$ and $T_P$. Filled circles -- 9.8-mm straw, 
open circles -- 18-mm straw. Curves: formula (\ref{tpfitform}) for the plot A and 
(\ref{frequency}) for B; solid lines -- 9.8-mm straw, dashed lines -- 18-mm straw.
}
\label{fig:coplot}
\end{figure}

The results of the lowest frequency measurements are shown in Figure \ref{fig:coplot}B 
along with the corresponding calculation results based on (\ref{frequency}). The 
error bars show the frequency uncertainty of 1 Hz defined by the width of the observed 
spectrum peaks.

The prediction (\ref{frequency}) for the 9.8-mm straw satisfactorily describes the measurement results 
in the vicinity of NA62 design parameters ($T_0 = 1500$~gf, $P = 1$~ at). However, the overall
discrepancy reaches 2 Hz, which is more than the measurement precision.
Nevertheless, the obtained qualitative description of the straw vibration confirms that 
the lowest frequency of the straw is significantly diminished by the internal overpressure in spite of 
the increase in straw wall stress according to (\ref{tformula}).

\section{Minimum straw pre-tension evaluation}

We can use the obtained Poisson's ratio to evaluate the minimum straw 
pre-tension $T_0$ in the straw-based NA62 spectrometer operating in vacuum \cite{NA62TD}.

There are two requirements for $T_0$. The buckling limit $T_P > 0$ (where $T_P$ is 
calculated from (\ref{tpfitform}) for $P = 1$~at) is defined by the Poisson's ratio 
and the straw radius. The flexural rigidity contribution (\ref{fcrit}) to the buckling prevention 
is small for the long NA62 straw and may be included into the safety margin. Moreover, 
the experimental design requirements define the gravitational deviation (sagitta) limit for 
a horizontal straw. Sagitta $Sag = y_s(0)$ calculated from (\ref{static}) depends on the 
effective tension, flexural rigidity, and straw length.

The sagitta requirement $Sag < 100$~$\mu$m of NA62 \cite{NA62TD} cannot be satisfied using a reasonable pre-tension for the complete 2.1-m straw. Hence, special supporting spacers 
dividing each straw into three equal parts are implemented in the NA62 spectrometer \cite{NA62TD}. 
These spacers fix a straw only in one of the lateral directions, and thus, they do not change 
the small buckling critical load value (\ref{fcrit}) ignored in this study.

Straw symmetry near each spacer ensures that the boundary conditions of (\ref{static}) 
are satisfied. Hence, we use this solution with $L=0.7$~m in order to calculate the sagitta.

\begin{figure}[t]
\begin{center}
\resizebox{\columnwidth}{!}{%
\setlength{\unitlength}{1mm}
\begin{picture}(166,80)
\includegraphics[width=80mm]{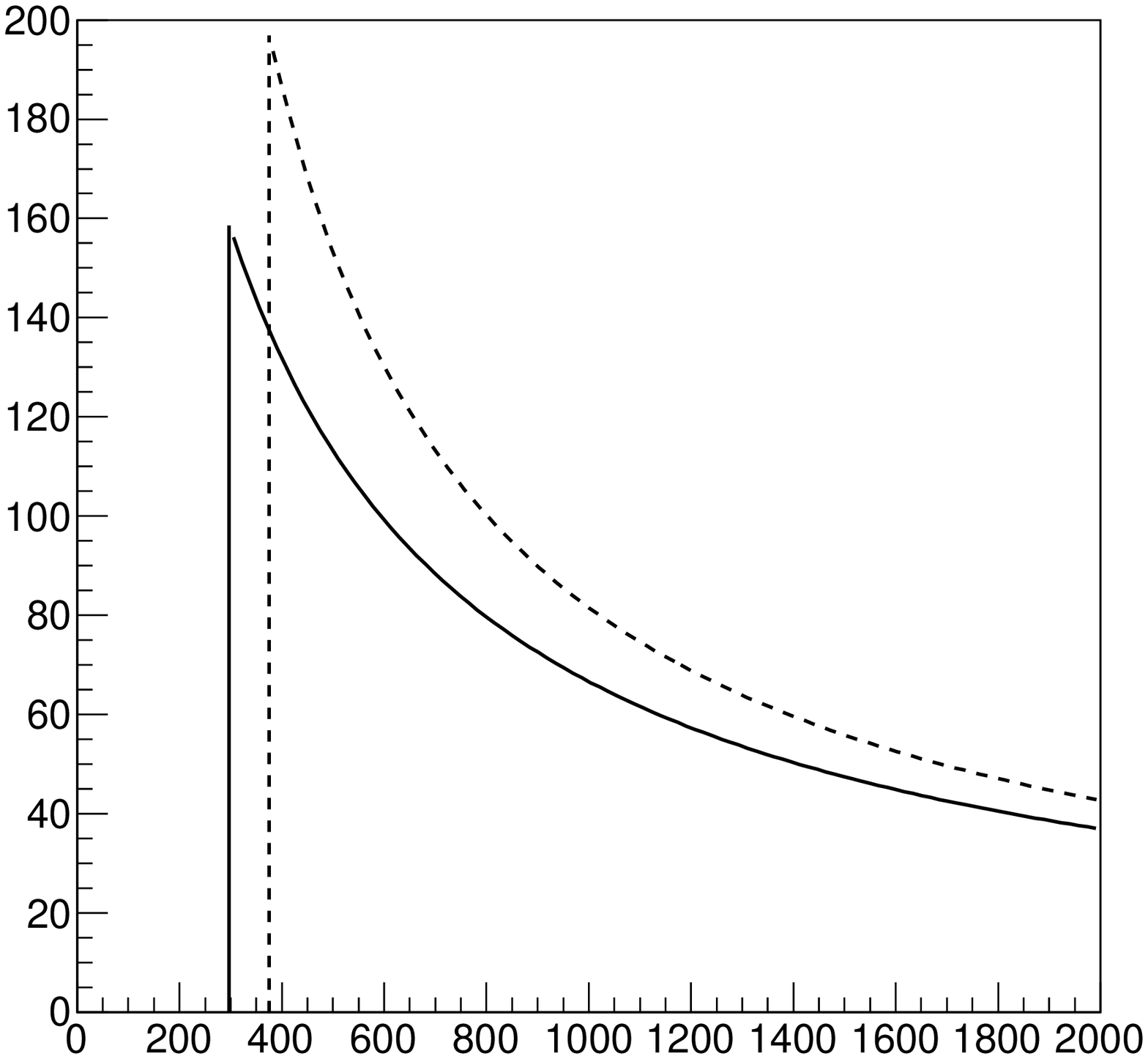}
\includegraphics[width=80mm]{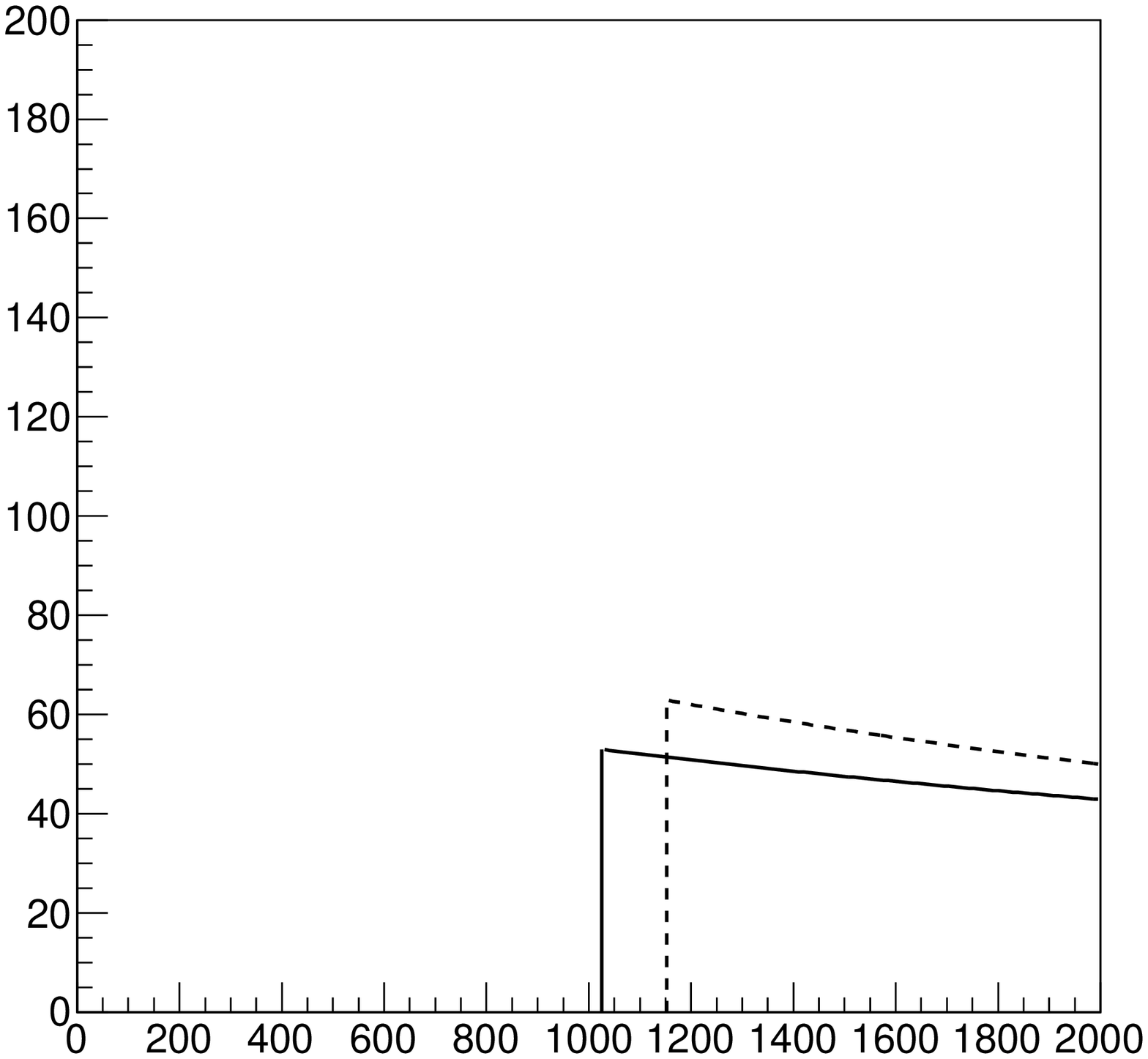}

\put(-95,62){A}
\put(-15,62){B}
\put(-45,0){$T_0$, gf}
\put(-125,0){$T_0$, gf}
\put(-82,74){$Sag$, $\mu$m}
\put(-162,74){$Sag$, $\mu$m}
\end{picture}
}\end{center}
\caption{Sagitta in vacuum $Sag$ versus pre-tension $T_0$ of 70-cm-long straw segment for 
9.8-mm straw (A) and 18-mm straw (B). Vertical lines mark the buckling limit $T_P=0$. Solid 
line is calculated for the central values for all the parameters. Dashed lines show the worst 
case (see text).
}
\label{fig:sagitta}
\end{figure}

The results of the sagitta calculation and buckling limit estimations are shown in Figure \ref{fig:sagitta}A 
for the 9.8-mm straw in the NA62 spectrometer. The corresponding results for the 18-mm straw, assuming the same 
straw length of 70 cm, are shown in Figure \ref{fig:sagitta}B for comparison.

Apart from the most probable sagitta values and buckling limits, the worst cases are also shown 
in Figure \ref{fig:sagitta}. The worst case for each straw corresponds to $\nu$ (\ref{nuform}) 
central value diminished by the tripled total uncertainty evaluated for each $T_0$ from the 
measured $\mu_\parallel$ and $k$ considering their correlation. All other parameters 
for this worst-case scenario are considered at their uncertainty limits leading to the largest sagitta 
and the easiest buckling.

It can be observed from Figure \ref{fig:sagitta} that, for the NA62 spectrometer, a straw pre-tension value above 
900 gf guarantees $Sag < 100$~$\mu$m (NA62 requirement), whereas the worst case of the buckling limit 
is below 400 gf. Thus, both the NA62 nominal straw pre-tension of $1.5$~kgf and the factual 
minimum pre-tension ($\approx 1.2$~kgf) \cite{NA62TD} have a good safety margin.

However, if the 18-mm straw is used in the same detector design, the minimum $T_0$ would be 
defined by the increased buckling limit above $1.15$~kgf with a small safety margin, whereas 
the gravitational sagitta will be always below 100~$\mu$m for the pre-tension above the 
buckling limit.

\section*{Conclusions}

A new technique for the study of mechanical properties of straw tubes subjected to inner pressure and 
longitudinal tension has been tested using a specially built test bench. It includes the measurement
of Poisson's ratio of a straw wall, which defines the buckling limit of a straw with a given radius under
a definite inner overpressure.

The axial Poisson's ratio $\mu_\parallel$ for Hostaphan\textsuperscript{\textregistered} foil is measured 
for two specimens under the conditions close to those of a detector operating in vacuum. The lateral 
Poisson's ratio $\mu_\perp$ is evaluated using the elasticity moduli in two directions provided by the 
foil producer. The spectra measurements of straw oscillations qualitatively confirm the effective 
tension predictions based on the measured $\mu_\parallel$.

The minimum pre-tension requirement for the NA62 spectrometer is re-evaluated based on 
the measurement results. The obtained limit confirms the detector design pre-tension with a 
safety margin of approximately 600 gf. The tested technique can be used for the development of future straw trackers.

\bibliography{pressure}

\end{document}